\documentclass[aps,prl,citeautoscript,reprint,superscriptaddress,floatfix,footinbib,showkeys]{revtex4-2}
\usepackage{setspace}
\usepackage[utf8]{inputenc}
\usepackage[colorlinks=true,linkcolor=black,urlcolor=black,filecolor=black,citecolor=black]{hyperref}
\usepackage{color,soul}
\usepackage{amsmath}
\usepackage{amsfonts}
\usepackage{amssymb}
\usepackage{graphicx}
\usepackage{dcolumn}
\usepackage{bm}
\usepackage{subfigure}
\usepackage{booktabs}

\begin{document}
\title{Breakdown of the Wiedemann-Franz law in an interacting quantum Hall metamaterial}
\author{Patrice Roche} 
\affiliation{Université Paris-Saclay, CEA, CNRS, SPEC, 91191 Gif-sur-Yvette, France} 
\author{Carles Altimiras} \affiliation{Université Paris-Saclay, CEA, CNRS, SPEC, 91191 Gif-sur-Yvette, France} \author{François D. Parmentier} \affiliation{Université Paris-Saclay, CEA, CNRS, SPEC, 91191 Gif-sur-Yvette, France} \affiliation{Laboratoire de Physique de l’Ecole normale sup\'erieure, ENS, Universit\'e PSL, CNRS, Sorbonne Universit\'e, Universit\'e Paris Cit\'e, F-75005 Paris, France } 
\author{Olivier Maillet}\email{olivier.maillet@cea.fr}
\affiliation{Université Paris-Saclay, CEA, CNRS, SPEC, 91191 Gif-sur-Yvette, France}

\begin{abstract}
Coulomb interactions deeply affect quantum transport in simple ballistic systems, but their impact on scaled up ballistic structures remains underexplored. Here we theoretically consider a chain of small metallic dots with frozen charge dynamics connected by ballistic channels. We identify a neutral mode of transport that is specific to a chain with at least two islands and entwines local diffusion by neutral excitations with long-range correlations between the islands' charge states. We show, as an experimentally measurable signature of this many-body behavior, that the Wiedemann-Franz law is violated with a Lorenz ratio scaling as the square root of the chain's length. 
\end{abstract} 

\maketitle

The simple combination of ballistic electron transport with a small electron reservoir, featuring strong on-site Coulomb repulsion, entails a wide variety of many-body phenomena. These include e.g.
quantum criticality in the multichannel Kondo effect \cite{IftikharNature2015,IftikharScience2018}, Tomonaga-Luttinger liquid physics \cite{AnthorePRX2018}, electron ``teleportation" \cite{DuprezScience2019}, charge and statistics fractionalization \cite{LeePRL2020,MorelPRB2024}, and the suppression of exactly one ballistic channel for the heat flow, coined "Heat Coulomb Blockade" (HCB) \cite{SlobodeniukPRB2013,SivreNatPhys2018,SpanslattPRB2024}. The "Coulomb island" in these realizations may still be considered a fermion reservoir as it contains $\sim 10^9$ electrons, while its very small stray capacitance $C$ (typically $2-3$ fF) strongly impacts the dynamics of collective electromagnetic degrees of freedom such as the island's macroscopic charge \cite{SlobodeniukPRB2013,StablerPRB2023,MorelPRB2024}. 

One may wonder about the emergence of new collective behaviours when scaling up to a chain of islands. Aside from a few theoretical works \cite{KarkiPRB2022,KarkiPRL2023,KiselevPRB2023,KarkiPRB2024,KarkiPRB2025} and one recent experimental tour-de-force \cite{PouseNatPhys2023} investigating fractionalization and the Kondo effect on a two-site chain, one paper only \cite{StäblerPRR2024} considered, in detail, the extension to an array. In most of these works, many-body physics was probed via the scattering properties of a quantum point contacts. Yet, many-body effects are still present even for entirely ballistic transport. Furthermore, they can be observed directly with thermal transport \cite{JezouinScience2013,SrivastavSciAdv2019,DuttaScience2022}, which, contrarily to charge transport, allows revealing charge-neutral excitations \cite{LeBretonPRL2022}. A strong marker of those effects is whether the Wiedemann-Franz (WF) law is satisfied: a departure of the Lorenz ratio between heat and charge conductance from $\mathcal{L}_0=\pi^2k_\mathrm{B}^2/3e^2\approx2.44\times10^{-8}\,\mathrm{V}^2/\mathrm{K}^2$ signals strong interactions \cite{KubalaPRL2008,CrossnoScience2016,DuttaPRL2017,BanerjeeNature2017,BanerjeeNature2018}. Here we show that thermal transport through a 1D array of Coulomb islands linked by ballistic channels is governed by a non-trivial interplay of the charge mode and neutral modes through the array's temperature field. This interplay is heralded by a violation of WF law opposite to that of HCB for a single island: heat becomes more efficiently transported than electricity. We pinpoint the role played by qualitatively different heat relaxation channels at a single island level, and derive the scaling laws governing heat flow when increasing chain size. Our approach generalizes the work of Stäbler et al. \cite{StäblerPRR2024} to multiple edge channels, giving rise to new physics in a minimal experimental arrangement that notably does not involve partitioning by a quantum point contact, facilitating experimental implementation.
Note that the ballistic edge channels of the integer quantum Hall regime are used as an illustration throughout the Letter, but our results apply more generally to any type of system featuring chiral ballistic channels \cite{KonigScience2007,ChangRMP2022}.
\begin{figure*}[ht!]
	\centering
	\includegraphics[width=1\textwidth]{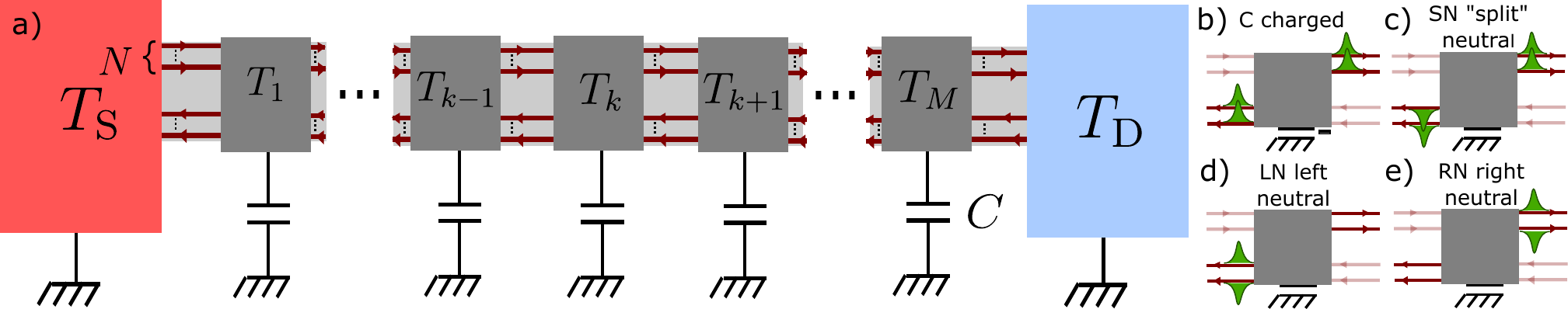}
	\caption{a) A chain of $M$ identical metallic nodes connecting two large reservoirs having different temperatures $T_\mathrm{S},\,T_\mathrm{D}$. Each metallic node is galvanically connected to its nearest neighbor via $N$ chiral ballistic edge states, and possesses a capacitance to the ground $C$. b)-e) Sketch representations of the four current noise transmission modes in the HCB regime for $N=2$: b) charged mode (C), where a net charge leaves the island, which modifies the stored charge on the capacitor (charge accumulation). c) "Split neutral" (SN) mode, where net charges with equal but opposite values simultaneously leave the island on each side through each channel, leaving the island's net charge unchanged. d) Left and e) right neutral (LN and RN) modes, where charges with zero sum leave the island on the same side, again conserving the island's net charge.}\label{fig_HCB_setup}
\end{figure*}

\textit{Description of the system - }The chain of $M$ Coulomb islands, connected to source and drain reservoirs of respective temperatures $T_\mathrm{S}$ and $T_\mathrm{D}$ by its ends, is shown in Fig. \ref{fig_HCB_setup}a. We first write the conservation of charge $q_k$ for each island $k$ and $N$ edge channels, in a Langevin approach \cite{SlobodeniukPRB2013}:
\begin{equation}
\begin{split}
    \label{charge_conservation}
    \frac{\mathrm{d}q_k}{\mathrm{d}t}=\sum_{\ell=1}^N\left(\delta I^{T_{k-1}}_{\mathrm{R},\ell}+\delta I^{T_{k+1}}_{\mathrm{L},\ell}-\delta I^{T_k}_{\mathrm{R},\ell}-\delta I^{T_k}_{\mathrm{L},\ell}\right)\\+NG_0(\delta V_{k-1}-2\delta V_k+\delta V_{k+1}),
\end{split}
\end{equation}
where the $\delta I_{\mathrm{L/R},\ell}^{T_k}$ terms are Johnson-Nyquist current noises emitted by the islands at thermal equilibrium temperature $T_k$. $\delta V_k$ is the fluctuating potential for island $k$, linked to the island charge $q_k$ via $\delta V_k=q_k/C$, where we assume the same capacitance $C$ for each island. Two regimes in Eq. (\ref{charge_conservation}) can be distinguished: first, if the island's capacitance $C$ is large enough, incoming current fluctuations at the typical thermal frequencies $\sim k_BT/\hbar\gg G_0/C$ flow to the ground through the capacitance. The island thus absorbs all incoming thermal radiation like an ideal black body, then emitting Johnson-Nyquist current noise in all channels (in an uncorrelated way) with equal average power $\pi^2k_\mathrm{B}^2T_k^2/6h$. In that situation voltage fluctuations vanish (each island is effectively grounded at the considered frequencies) and the (squared) temperature profile is that of a diffusive medium in a discretized version, independent of $N$, i.e.:
\begin{equation}
\left(T_k^\mathrm{diff}\right)^2=\left(1-\frac{k}{M+1}\right)T_\mathrm{S}^2+\frac{k}{M+1}T_\mathrm{D}^2,
    \label{T_sq_noninteracting}
\end{equation}
with a corresponding source-drain heat flow $\dot{Q}^{\mathrm{diff}}=N\kappa_0\left(T_\mathrm{S}^2-T_\mathrm{D}^2\right)/(M+1)$. Note that this may be obtained also through a standard Landauer approach for heat (see End Matter) \cite{BlencowePRA2000}. In the opposite case $k_BT/\hbar\ll G_0/C$, in the so-called "Heat Coulomb Blockade" (HCB) regime, charge cannot vary on the island. Therefore $\mathrm{d}q_k/\mathrm{d}t= 0$ in Eq. (\ref{charge_conservation}): thermal current noise cannot flow through the capacitance, which correlates incoming and outgoing physical ballistic channels that are no longer independent from each other. In that limit, it is convenient to consider outgoing charge propagation modes according to the following decomposition: $2N-1$ neutral modes carry energy but no net charge, while one remaining charge mode corresponds to simultaneous emission of charges equally in each physical ballistic channel \cite{SlobodeniukPRB2013,MorelPRB2024}. 

\textit{Power balance in the HCB regime - }We can detail this decomposition for channels going out of an island, as sketched in Fig. \ref{fig_HCB_setup}b)-e) (see Supp. Mat. \cite{SM} for the explicit decomposition of bosonic fields). This decomposition is not unique \cite{MorelPRB2024}, but is intuitive in the HCB regime. A charge (C) mode of Fig. \ref{fig_HCB_setup}b changes the net overall charge of an island, and thus, in the HCB regime, can be triggered only by another net charge (which needs not be quantized) incoming on the island. In other words, thermal fluctuations carried by the C mode cannot be spontaneously generated when $k_\mathrm{B}T\ll \hbar G_0/C$, as the charge on the capacitor is frozen. Random emission can occur through the "split" neutral (SN, Fig. \ref{fig_HCB_setup}c) mode, as it consists in the emission of $2N$ packets, $N$ in each direction with equal charges but opposite signs (thus preserving the island's charge). Therefore, net charges do arrive to the adjacent islands. 
Since they cannot be accommodated here, their arrival triggers the emission of the neighbor C modes to ensure charge neutrality, which then cascades down the chain. Thus, no energy is dissipated in the other islands with this process, and all the power emitted by SN modes can be dissipated only in the source and drain. As such, the SN mode is a cooling pathway of the island directly toward the source and drain, with respective amounts determined by the number of islands on the way, acting as 50:50 power splitters. The remaining $2N-2$ neutral modes are the $N-1$ left (LN, Fig. \ref{fig_HCB_setup}d) and $N-1$ right (RN, Fig. \ref{fig_HCB_setup}e) ones which can be spontaneously emitted on a single side and simply absorbed by the neighbor islands in a diffusion-like process, just like in the non-interacting case. They are a cooling pathway toward the neighbor islands which are then heated up, unlike with the SN mode. 

Since all power is dissipated in the source and drain in the cooling of all islands via the SN mode, this process tends to equalize all temperatures. Meanwhile, diffusion via LN/RN modes makes them closer to the profile of Eq. (\ref{T_sq_noninteracting}). To obtain each island's temperature and the heat flowing from source to drain resulting from this balance between interactions and diffusion, we first invert the system of equations (\ref{charge_conservation}) for each island $k$:
\begin{equation}
\begin{split}
    \label{voltage_island}
\delta V_k=\frac{M+1-k}{G_0N(M+1)}\sum_{j< k}\sum_{\ell=1}^N\left(\delta I^{T_j}_{\mathrm{R},\ell}-\delta I^{T_j}_{\mathrm{L},\ell}\right)\\+\frac{k}{G_0N(M+1)}\sum_{j> k}\sum_{\ell=1}^N\left(\delta I^{T_j}_{\mathrm{L},\ell}-\delta I^{T_j}_{\mathrm{R},\ell}\right)\\+\sum_{\ell=1}^N\frac{\left[(M+1-k)\left(\delta I^{T_\mathrm{S}}_{\ell}-\delta I^{T_k}_{\mathrm{L},\ell}\right)+k\left(\delta I^{T_\mathrm{D}}_{\ell}-\delta I^{T_k}_{\mathrm{R},\ell}\right)\right]}{G_0N(M+1)}.
\end{split}
\end{equation}
We then write the power emitted in any outgoing channel $\ell$, $P_\mathrm{\gamma\in\{\mathrm{L,R}\},\ell}^k=\int_{-\infty}^{+\infty} S_{I_{\gamma,\ell}^{k}}(\omega)\mathrm{d}\omega/4\pi G_0$, where the outgoing current noise writes $\delta I_{\gamma,\ell}^k=\delta I_{\gamma,\ell}^{T_k}+G_0\delta V_k$, with a noise spectral density $2\pi S_{I_{\gamma,\ell}^k}(\omega)\delta(\omega+\omega')=\left\langle\delta I_{\gamma,\ell}^k(\omega)\delta I_{\gamma,\ell}^k(\omega')\right\rangle$, and equilibrium thermal noise $S_{I^{T_k}_{\gamma,\ell}}(\omega)=G_0\hbar\omega\left[1/\left(1-e^{-\hbar\omega/k_\mathrm{B}T_k}\right)-\Theta(\omega)\right]$. Strikingly, an asymmetry appears: $P_{\mathrm{L},\ell}^k\neq P_{\mathrm{R},\ell}^k$, which boils down to the voltage/current cross-correlation: indeed $\left\langle\delta I_{\mathrm{L},\ell}^{T_k}(-\omega)\delta V_k(\omega)\right\rangle\neq\left\langle\delta I_{\mathrm{R},\ell}^{T_k}(-\omega)\delta V_k(\omega)\right\rangle$. This asymmetry occurs because, from one island's perspective, the emitted current noise is fed back into the island. The feedback weight is determined by the number of islands toward the source and the number toward the drain, which in the general case differ.
\begin{figure}
	\centering
\includegraphics[width=0.48\textwidth]{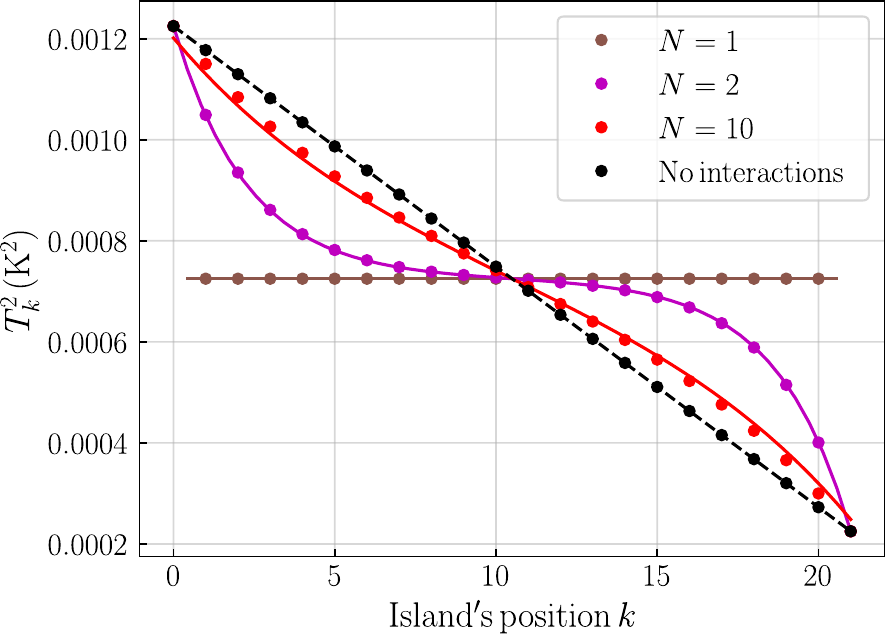}
	\caption{Squared temperature profile along the chain, for $M=20$ islands, and source and drain temperatures $T_\mathrm{S}=35$ mK, $T_\mathrm{D}=15$ mK, for three channel numbers $N=1,2,10$. Solid lines are the application of Eq. (\ref{T_profile_simple}) derived in the long chain limit $M\gg\Delta$. The dashed line corresponds to a high temperature regime $k_BT\gg\hbar G_0/C$ limit, for which the diffusive profile of Eq. (\ref{T_sq_noninteracting}) is recovered.}\label{T_profile}
\end{figure}
The total power $P_\mathrm{\mathrm{L}}^k=\sum_\ell P_{\mathrm{L},\ell}^k$ emitted in the left channels is:
\begin{equation}
\begin{split}
P_\mathrm{L}^k/\kappa_0=\left[N-1+\frac{2k^2}{(M+1)^2}\right]T_k^2\\+\frac{(M+1-k)^2T_\mathrm{S}^2+k^2T_\mathrm{D}^2}{(M+1)^2}\\+2\frac{(M+1-k)^2\sum_{j=1}^{k-1}T_j^2+k^2\sum_{j=k+1}^{M}T_j^2}{(M+1)^2},
\end{split}
    \label{power_left}
\end{equation}
with $\kappa_0=\pi^2k_\mathrm{B}^2/6h$. The contribution of the $(N-1)$ LN modes amounts to $(N-1)\kappa_0T_k^2$, while the rest is made of SN/C contributions that involve all temperatures (see Supplemental Material \cite{SM}). Likewise, the power radiated to the right is:
\begin{equation}
\label{power_right}
P_\mathrm{R}^k=P_\mathrm{L}^k+\frac{2(M+1-2k)}{M+1}\kappa_0T_k^2,    
\end{equation}
where we highlight the power emission asymmetry, which vanishes for $k\approx(M+1)/2$, for an island in the middle of the chain where electromagnetic environments are symmetric, which results in equal feedback from both sides. Note that we recover, for $N=1$, an emitted power scaling linearly with the chain's size $M$ in the middle of the chain in the large chain limit, just like the "open BC case" of Ref. \cite{StäblerPRR2024}. 

\textit{Temperature profile - }The temperature profile of the chain is obtained from the power balance for each island, neglecting electron-phonon heat flow: $P_\mathrm{L}^k+P_\mathrm{R}^k=P_\mathrm{L}^{k+1}+P_\mathrm{R}^{k-1}$, with boundary conditions $P_\mathrm{R}^0=N\kappa_0T_\mathrm{S}^2,\,P_\mathrm{L}^{M+1}=N\kappa_{0}T_\mathrm{D}^2$ (source/drain are assumed to be ideal, grounded reservoirs). We obtain the following diffusion equation:
\begin{equation}
    \label{Heat_diffusion}
(N-1)\left[T_{k+1}^2-2T_{k}^2+T_{k-1}^2\right]=4\frac{T_k^2-T_\mathrm{mid}^2}{M+1},
\end{equation}
with $T_\mathrm{mid}^2=(T_\mathrm{S}^2+T_\mathrm{D}^2)/2$. The left-hand side term, akin to a discrete Laplacian, corresponds to the diffusive contribution of the LN/RN neutral modes. As these neutral excitations are fully absorbed by the neighbor metallic islands, they contribute only locally to the chain thermalization, from a given island to its first neighbors, and appear identically in a non-interacting situation. The right-hand side term, on the contrary, is absent without interactions. It reflects the fact that the SN mode emission is energetically equivalent for all islands. This tends to equalize temperatures and leads to a large deviation from the diffusive temperature profile of Eq. (\ref{T_sq_noninteracting}). The competition between the two dynamics in Eq. (\ref{Heat_diffusion}) is highlighted by a characteristic "decay length" $\Delta=\sqrt{(N-1)(M+1)}/2$, which depends on the chain size.
\begin{figure}
	\centering
\includegraphics[width=0.48\textwidth]{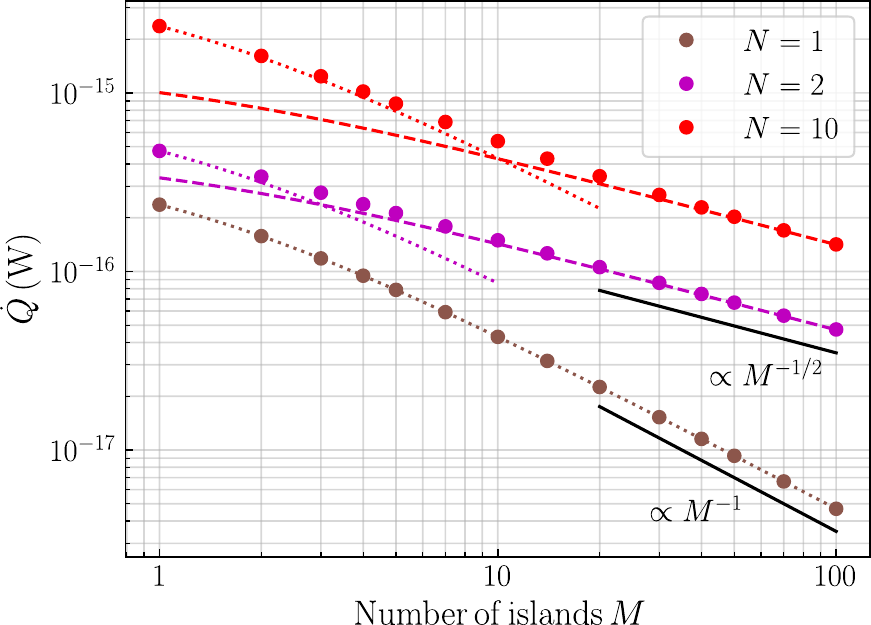}
	\caption{Source-drain heat flow as a function of the chain length, for $N=1$ (brown), 2 (magenta) and 10 (red) channels, $T_\mathrm{S}=
    35$ mK and $T_\mathrm{D}=15$ mK. Dots represent exact heat flows obtained by Eq. (\ref{Qdot_local_nonlocal}). Dashed lines are the approximations obtained in the long chain limit $M\gg \Delta$, i.e. the application of Eq. (\ref{Qdot_0_large_chain}). Dotted lines are the diffusive heat flows, obtained in the absence of interactions (see text).}\label{heat_flow}
\end{figure}
\begin{figure}[ht!]
	\centering
\includegraphics[width=0.48\textwidth]{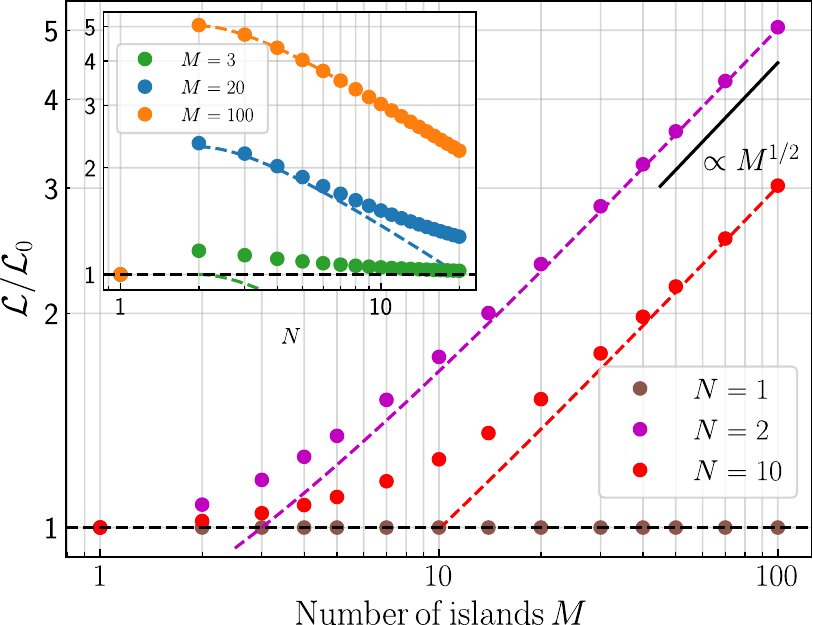}
	\caption{a) Lorenz ratio as a function of the chain size, for $N=1,\,2,\,10$. Dashed magenta and red lines are the application of Eq. (\ref{Lorenz_ratio}) for the corresponding $N$. The black dashed line follows the WF law. Inset: Lorenz ratio as a function of the number of edge channels $N$, for chain sizes $M=3,\,20,\,100$. Dashed colored lines are the corresponding applications of Eq. (\ref{Lorenz_ratio}).}\label{fig_Lorenz}
\end{figure}
The temperature profile satisfying the heat diffusion equation, Eq. (\ref{Heat_diffusion}), has a complicated but analytical form (see End Matter), with exponential features on each side of the chain: diffusive dynamics is washed out within a distance $\sim\Delta$ of both the source and drain reservoirs, which initiate thermalization. An approximate, simpler version is obtained in the limit $M\gg \Delta$ (or equivalently $M\gg N$):
\begin{equation}
    \label{T_profile_simple}
T_k^2\underset{M\gg \Delta}{\approx}T_\mathrm{mid}^2+\frac{T_\mathrm{S}^2-T_\mathrm{D}^2}{2}\left[e^{-k/\Delta}-e^{-(M+1-k)/\Delta}\right]
\end{equation}

Remarkably, the decay length $\Delta$ as well as the diffusive term in Eq. (\ref{Heat_diffusion}) vanish for $N=1$, where each island $k$ has a temperature $T_\mathrm{mid}$. Intuitively, this is explained by the fact that neutral excitations cannot be emitted by the source (drain) when there is a single outgoing channel toward island 1 ($M$). Therefore, the chain is thermally decoupled from the source and drain reservoirs. Inside the chain, LN and RN modes do not exist for $N=1$, and only the C and SN modes survive. Since the C mode spontaneous excitation is forbidden in the HCB regime, the only cooling pathway is the SN mode for every island alike: therefore, they all have the same temperature.

\textit{Heat flow and Lorenz ratio - }The squared temperature profile across the chain, obtained with the full solution of Eq. (\ref{Heat_diffusion}) is shown in Figure \ref{T_profile} for $M=20$ islands. A sharp drop in temperature is indeed present in the $N=1$ channel limit, due to the absence of neutral modes mediating heat diffusion as explained above. The agreement with the long chain approximation is good when the number of channels is small (here $N=1,\,2$), but breaks down for a large number of channels ($N=10$ here). In that limit, the diffusive solution, Eq. (\ref{T_sq_noninteracting}), is close to the obtained profile. This is explained by the fact that in this limit, heat diffusion by the $2\times(N-1)$ RN and LN modes is comparatively much larger than the contribution due to the SN mode. 

We then write the global heat flow, which, by continuity and in absence of electron-phonon cooling, is the net heat flow between any two given successive islands, $\dot{Q}=P_\mathrm{L}^{k+1}-P_\mathrm{R}^{k}$. It is composed of the local diffusion term, but also has a \textit{non-local} dependence on all the islands' temperatures: 
\begin{equation}
\begin{split}
   \dot{Q}/\kappa_0=(N-1)\left(T_{k}^2-T_{k+1}^2\right)\\+\frac{T_\mathrm{S}^2-T_\mathrm{D}^2}{M+1}+\frac{4\sum_{j=1}^k\left(T_j^2-T_\mathrm{mid}^2\right)}{M+1}.    
\end{split}
    \label{Qdot_local_nonlocal}    
\end{equation}
In the singular case $N=1$ the heat flow retains its non-local contribution only due to the SN mode. Since all sites have the same temperature $T_\mathrm{mid}$, the dependence on the sites' temperature vanishes, and we have $\dot{Q}=\kappa_0\left(T_\mathrm{S}^2-T_\mathrm{D}^2\right)/(M+1)$, as shown in Fig. \ref{heat_flow}: heat does flow from source to drain despite the chain's thermal decoupling. This expression is identical to the non-interacting one, which appears counterintuitive since here the local diffusive contribution vanishes. This is the consequence of successive power splittings by the islands, rerouting most of the heat flow emitted by the source toward it. When adding neutral modes, a given fraction of the heat flow does not undergo such redistribution and is instead diffused, leading to more power actually transmitted toward the drain. This appears in the power emission asymmetry in Eq. (\ref{power_right}), which does depend on $N$ indirectly through the temperature: on the left part of the chain, the asymmetry is enhanced towards the drain for $N\geq 2$ since $T_k>T_\mathrm{mid}$ over a distance $\Delta\sim(M+1)^{1/2}$; meanwhile, it is reduced toward the source on the right part of the chain because $T_k<T_\mathrm{mid}$ over $\Delta$. 
One can write the heat flow between the source and the first island in the multichannel case $N\geq 2$, using the continuity of the heat flow and the temperature profile in the large chain limit $M\gg N$. Since $T_1^2\approx T_\mathrm{S}^2-\left(T_\mathrm{S}^2-T_\mathrm{D}^2\right)/2\Delta$, we obtain at lowest order,
\begin{equation}
    \label{Qdot_0_large_chain}
    \dot{Q}/\kappa_0\underset{M\gg\Delta}{\approx}\sqrt{\frac{N-1}{M+1}}\left(T_\mathrm{S}^2-T_\mathrm{D}^2\right).
\end{equation}
In the large chain limit $\dot{Q}\sim M^{-1/2}\left(T_\mathrm{S}^2-T_\mathrm{D}^2\right)$, as shown in Fig. \ref{heat_flow}, which is in contrast with a diffusive-type heat flow and the $N=1$ case, where $\dot{Q}$ scales as $M^{-1}\left(T_\mathrm{S}^2-T_\mathrm{D}^2\right)$. Meanwhile, the chain's DC conductance $G_\mathrm{el}=NG_0/(M+1)$ is unaffected by interactions. Therefore, for $N>1$, Wiedemann-Franz law fails, with a Lorenz ratio
\begin{equation}
\mathcal{L} =\frac{\dot{Q}/(T_\mathrm{S}-T_\mathrm{D})}{G_\mathrm{el}(T_\mathrm{S}+T_\mathrm{D})/2} \underset{M\gg\Delta}{\approx}\frac{\sqrt{(N-1)(M+1)}}{N}\mathcal{L}_0,
\label{Lorenz_ratio}
\end{equation}
that scales as $\sqrt{M}$: a long chain is a better electrical insulator than a thermal one, and it may be thought of as a heat sink. In Figure \ref{fig_Lorenz}a) we show the Lorenz ratio evolution with the chain size for $N=1,2$ and 10 channels. For $N=1$, the ratio remains equal to 1, since the heat flow matches the diffusive solution. For $N\neq1$, the Lorenz ratio indeed grows as $M^{1/2}$, provided that $M\gg N$. We also recover that the $M=1$ is a trivial limit case, as any feedback is suppressed by the reservoirs. In the inset we represent the Lorenz ratio as a function of the number of channels for $M=3,20$ and 100 islands. As pointed out above, the $N=1$ case is singular. For $N\geq2$ substantial deviations to the WF law occur at only a low number of channels $N\ll M$. At larger $N$ diffusive dynamics takes over and the ratio converges to $\mathcal{L}_0$ again, while the scaling law of Eq. (\ref{Lorenz_ratio}), valid only in the large chain limit $M\gg\Delta$, is no longer accurate (see green dashed line in Fig. \ref{fig_Lorenz}b). 

\textit{Discussion} - Experimentally, the computed heat flows for reasonable configurations ($M\lesssim 30$) lie in the 0.01-1 fW range, which is challenging but realistically measurable \cite{JezouinScience2013,SivreNatPhys2018,LeBretonPRL2022}. Note that the bigger $N$, the more relaxed is the condition on full HCB. For $N=2$, this condition is well achieved for temperatures smaller than 80 mK, with capacitances $C\sim 3$ fF, which means that a spread over individual islands' capacitances due, e.g., to standard inhomogeneities in a fabrication process, is tolerable as long as each capacitance well exceeds $k_\mathrm{B}T_\mathrm{S}/Ne^2$. Importantly, we neglect here the finite propagation time of current fluctuations when addressing feedback, which is considered instantaneous. Assuming that current fluctuations propagate at a plasmon velocity $v\sim10^5$ m/s, somewhat intermediate between gated GaAs-based 2DEGs \cite{KumadaPRB2011} and graphene \cite{PetkovićJPhysD2014}, this is valid for a sample's physical length $L\ll  hv/k_BT\sim120\,\mu$m at 50 mK. 

In summary, we have introduced a meta-material with strong on-site interactions, leading to non-trivial heat transport that is larger than the diffusive limit valid in the absence of interactions. Our calculations may be readily extended to fractional edges. New signatures may appear by generalizing our configuration, for instance to a 2D array, or by considering capacitive couplings directly between sites. An imbalance of counter-propagating channels may be engineered \cite{ZhangPRR2025} in order to study the impact of frozen charges on equilibration processes along edge states \cite{KanePRL1994,SpanslattPRB2024}. Finally, entropy signatures may also be revealed by adding quantum point contacts between islands \cite{KarkiPRB2022, hurvitz2025metallicislandarraysynthetic}, and could be measured with a charge detector \cite{HartmanNatPhys2018}.  
\section{Acknowledgments}
We thank A. Zhang, H. Duprez and D. C. Glattli for helpful discussions. 
O.M. acknowledges funding from the ANR (ANR-23-CE47-0002 CRAQUANT and ANR-25-ERCS-0006 ANYQUAC). F.D.P. was funded by the ERC (ERC-2018-STG \textit{QUAHQ}), by the “Investissements d’Avenir” LabEx PALM (ANR-10-LABX-0039-PALM), and by the Region Ile de France through the DIM QUANTIP.

\section{Competing interests}

The authors declare no competing interests.

\bibliography{KFP_simu}

@article{
JezouinScience2013,
author = {S. Jezouin  and F. D. Parmentier  and A. Anthore  and U. Gennser  and A. Cavanna  and Y. Jin  and F. Pierre },
title = {Quantum Limit of Heat Flow Across a Single Electronic Channel},
journal = {Science},
volume = {342},
number = {6158},
pages = {601-604},
year = {2013},
doi = {10.1126/science.1241912},
URL = {https://www.science.org/doi/abs/10.1126/science.1241912}}

@Article{SivreNatPhys2018,
author={Sivre, E.
and Anthore, A.
and Parmentier, F. D.
and Cavanna, A.
and Gennser, U.
and Ouerghi, A.
and Jin, Y.
and Pierre, F.},
title={Heat Coulomb blockade of one ballistic channel},
journal={Nature Physics},
year={2018},
month={Feb},
day={01},
volume={14},
number={2},
pages={145-148},
issn={1745-2481},
doi={10.1038/nphys4280},
url={https://doi.org/10.1038/nphys4280}
}

@article{SlobodeniukPRB2013,
  title = {Equilibration of quantum Hall edge states by an Ohmic contact},
  author = {Slobodeniuk, Artur O. and Levkivskyi, Ivan P. and Sukhorukov, Eugene V.},
  journal = {Phys. Rev. B},
  volume = {88},
  issue = {16},
  pages = {165307},
  numpages = {5},
  year = {2013},
  month = {Oct},
  publisher = {American Physical Society},
  doi = {10.1103/PhysRevB.88.165307},
  url = {https://link.aps.org/doi/10.1103/PhysRevB.88.165307}
}

@Article{BanerjeeNature2018,
author={Banerjee, Mitali
and Heiblum, Moty
and Umansky, Vladimir
and Feldman, Dima E.
and Oreg, Yuval
and Stern, Ady},
title={Observation of half-integer thermal Hall conductance},
journal={Nature},
year={2018},
month={Jul},
day={01},
volume={559},
number={7713},
pages={205-210},
issn={1476-4687},
doi={10.1038/s41586-018-0184-1},
url={https://doi.org/10.1038/s41586-018-0184-1}
}

@Article{BanerjeeNature2017,
author={Banerjee, Mitali
and Heiblum, Moty
and Rosenblatt, Amir
and Oreg, Yuval
and Feldman, Dima E.
and Stern, Ady
and Umansky, Vladimir},
title={Observed quantization of anyonic heat flow},
journal={Nature},
year={2017},
month={May},
day={01},
volume={545},
number={7652},
pages={75-79},
issn={1476-4687},
doi={10.1038/nature22052},
url={https://doi.org/10.1038/nature22052}
}

@article{DuttaScience2022,
author = {Bivas Dutta  and Vladimir Umansky  and Mitali Banerjee  and Moty Heiblum },
title = {Isolated ballistic non-abelian interface channel},
journal = {Science},
volume = {377},
number = {6611},
pages = {1198-1201},
year = {2022},
doi = {10.1126/science.abm6571},
URL = {https://www.science.org/doi/abs/10.1126/science.abm6571},
}

@article{AnthorePRX2018,
  title = {Circuit Quantum Simulation of a Tomonaga-Luttinger Liquid with an Impurity},
  author = {Anthore, A. and Iftikhar, Z. and Boulat, E. and Parmentier, F. D. and Cavanna, A. and Ouerghi, A. and Gennser, U. and Pierre, F.},
  journal = {Phys. Rev. X},
  volume = {8},
  issue = {3},
  pages = {031075},
  numpages = {14},
  year = {2018},
  month = {Sep},
  publisher = {American Physical Society},
  doi = {10.1103/PhysRevX.8.031075},
  url = {https://link.aps.org/doi/10.1103/PhysRevX.8.031075}
}

@Article{PouseNatPhys2023,
author={Pouse, Winston
and Peeters, Lucas
and Hsueh, Connie L.
and Gennser, Ulf
and Cavanna, Antonella
and Kastner, Marc A.
and Mitchell, Andrew K.
and Goldhaber-Gordon, David},
title={Quantum simulation of an exotic quantum critical point in a two-site charge Kondo circuit},
journal={Nature Physics},
year={2023},
month={Apr},
day={01},
volume={19},
number={4},
pages={492-499},
issn={1745-2481},
doi={10.1038/s41567-022-01905-4},
url={https://doi.org/10.1038/s41567-022-01905-4}
}

@article{LeBretonPRL2022,
  title = {Heat Equilibration of Integer and Fractional Quantum Hall Edge Modes in Graphene},
  author = {Le Breton, G. and Delagrange, R. and Hong, Y. and Garg, M. and Watanabe, K. and Taniguchi, T. and Ribeiro-Palau, R. and Roulleau, P. and Roche, P. and Parmentier, F. D.},
  journal = {Phys. Rev. Lett.},
  volume = {129},
  issue = {11},
  pages = {116803},
  numpages = {6},
  year = {2022},
  month = {Sep},
  publisher = {American Physical Society},
  doi = {10.1103/PhysRevLett.129.116803},
  url = {https://link.aps.org/doi/10.1103/PhysRevLett.129.116803}
}

@article{StäblerPRR2024,
  title = {Giant heat flux effect in nonchiral transmission lines},
  author = {St\"abler, Florian and Gadiaga, Alioune and Sukhorukov, Eugene V.},
  journal = {Phys. Rev. Res.},
  volume = {7},
  issue = {3},
  pages = {033228},
  numpages = {9},
  year = {2025},
  month = {Sep},
  publisher = {American Physical Society},
  doi = {10.1103/vw6r-2xcr},
  url = {https://link.aps.org/doi/10.1103/vw6r-2xcr}
}

@article{PetkovićJPhysD2014,
doi = {10.1088/0022-3727/47/9/094010},
url = {https://dx.doi.org/10.1088/0022-3727/47/9/094010},
year = {2014},
month = {feb},
publisher = {IOP Publishing},
volume = {47},
number = {9},
pages = {094010},
author = {Petković, Ivana and Williams, F I B and Glattli, D Christian},
title = {Edge magnetoplasmons in graphene},
journal = {Journal of Physics D: Applied Physics}
}

@article{KumadaPRB2011,
  title = {Edge magnetoplasmon transport in gated and ungated quantum Hall systems},
  author = {Kumada, N. and Kamata, H. and Fujisawa, T.},
  journal = {Phys. Rev. B},
  volume = {84},
  issue = {4},
  pages = {045314},
  numpages = {6},
  year = {2011},
  month = {Jul},
  publisher = {American Physical Society},
  doi = {10.1103/PhysRevB.84.045314},
  url = {https://link.aps.org/doi/10.1103/PhysRevB.84.045314}
}

@article{DuttaPRL2017,
  title = {Thermal Conductance of a Single-Electron Transistor},
  author = {Dutta, B. and Peltonen, J. T. and Antonenko, D. S. and Meschke, M. and Skvortsov, M. A. and Kubala, B. and K\"onig, J. and Winkelmann, C. B. and Courtois, H. and Pekola, J. P.},
  journal = {Phys. Rev. Lett.},
  volume = {119},
  issue = {7},
  pages = {077701},
  numpages = {5},
  year = {2017},
  month = {Aug},
  publisher = {American Physical Society},
  doi = {10.1103/PhysRevLett.119.077701},
  url = {https://link.aps.org/doi/10.1103/PhysRevLett.119.077701}
}

@article{KubalaPRL2008,
  title = {Violation of the Wiedemann-Franz Law in a Single-Electron Transistor},
  author = {Kubala, Bj\"orn and K\"onig, J\"urgen and Pekola, Jukka},
  journal = {Phys. Rev. Lett.},
  volume = {100},
  issue = {6},
  pages = {066801},
  numpages = {4},
  year = {2008},
  month = {Feb},
  publisher = {American Physical Society},
  doi = {10.1103/PhysRevLett.100.066801},
  url = {https://link.aps.org/doi/10.1103/PhysRevLett.100.066801}
}

@article{StablerPRB2023,
  title = {Mesoscopic heat multiplier and fractionalizer},
  author = {St\"abler, Florian and Sukhorukov, Eugene},
  journal = {Phys. Rev. B},
  volume = {108},
  issue = {23},
  pages = {235405},
  numpages = {12},
  year = {2023},
  month = {Dec},
  publisher = {American Physical Society},
  doi = {10.1103/PhysRevB.108.235405},
  url = {https://link.aps.org/doi/10.1103/PhysRevB.108.235405}
}

@article{SpanslattPRB2024,
  title = {Impact of potential and temperature fluctuations on charge and heat transport in quantum Hall edges in the heat Coulomb blockade regime},
  author = {Sp\aa{}nsl\"att, Christian and St\"abler, Florian and Sukhorukov, Eugene V. and Splettstoesser, Janine},
  journal = {Phys. Rev. B},
  volume = {110},
  issue = {7},
  pages = {075431},
  numpages = {24},
  year = {2024},
  month = {Aug},
  publisher = {American Physical Society},
  doi = {10.1103/PhysRevB.110.075431},
  url = {https://link.aps.org/doi/10.1103/PhysRevB.110.075431}
}

@article{MorelPRB2024,
  title = {Fractionalization and anyonic statistics in the integer quantum Hall collider},
  author = {Morel, Tom and Lee, June-Young M. and Sim, H.-S. and Mora, Christophe},
  journal = {Phys. Rev. B},
  volume = {105},
  issue = {7},
  pages = {075433},
  numpages = {14},
  year = {2022},
  month = {Feb},
  publisher = {American Physical Society},
  doi = {10.1103/PhysRevB.105.075433},
  url = {https://link.aps.org/doi/10.1103/PhysRevB.105.075433}
}

@article{
SrivastavSciAdv2019,
author = {Saurabh Kumar Srivastav  and Manas Ranjan Sahu  and K. Watanabe  and T. Taniguchi  and Sumilan Banerjee  and Anindya Das },
title = {Universal quantized thermal conductance in graphene},
journal = {Science Advances},
volume = {5},
number = {7},
pages = {eaaw5798},
year = {2019},
doi = {10.1126/sciadv.aaw5798},
URL = {https://www.science.org/doi/abs/10.1126/sciadv.aaw5798},
}

@article{BlencowePRA2000,
  title = {Universal quantum limits on single-channel information, entropy, and heat flow},
  author = {Blencowe, Miles P. and Vitelli, Vincenzo},
  journal = {Phys. Rev. A},
  volume = {62},
  issue = {5},
  pages = {052104},
  numpages = {8},
  year = {2000},
  month = {Oct},
  publisher = {American Physical Society},
  doi = {10.1103/PhysRevA.62.052104},
  url = {https://link.aps.org/doi/10.1103/PhysRevA.62.052104}
}

@article{ChangRMP2022,
  title = {Colloquium: Quantum anomalous Hall effect},
  author = {Chang, Cui-Zu and Liu, Chao-Xing and MacDonald, Allan H.},
  journal = {Rev. Mod. Phys.},
  volume = {95},
  issue = {1},
  pages = {011002},
  numpages = {33},
  year = {2023},
  month = {Jan},
  publisher = {American Physical Society},
  doi = {10.1103/RevModPhys.95.011002},
  url = {https://link.aps.org/doi/10.1103/RevModPhys.95.011002}
}

@article{
KonigScience2007,
author = {Markus König  and Steffen Wiedmann  and Christoph Brüne  and Andreas Roth  and Hartmut Buhmann  and Laurens W. Molenkamp  and Xiao-Liang Qi  and Shou-Cheng Zhang },
title = {Quantum Spin Hall Insulator State in HgTe Quantum Wells},
journal = {Science},
volume = {318},
number = {5851},
pages = {766-770},
year = {2007},
doi = {10.1126/science.1148047},
URL = {https://www.science.org/doi/abs/10.1126/science.1148047}}

@article{
DuprezScience2019,
author = {H. Duprez  and E. Sivre  and A. Anthore  and A. Aassime  and A. Cavanna  and U. Gennser  and F. Pierre },
title = {Transmitting the quantum state of electrons across a metallic island with Coulomb interaction},
journal = {Science},
volume = {366},
number = {6470},
pages = {1243-1247},
year = {2019},
doi = {10.1126/science.aaw7856},
URL = {https://www.science.org/doi/abs/10.1126/science.aaw7856}}

@misc{SM,
   title = {See Supplemental Materials at [URL] for details on bosonic charged/neutral modes construction and additional calculations},
}

@article{KarkiPRB2022,
  title = {Double-charge quantum island in the quasiballistic regime},
  author = {Karki, Deepak B. and Boulat, Edouard and Mora, Christophe},
  journal = {Phys. Rev. B},
  volume = {105},
  issue = {24},
  pages = {245418},
  numpages = {12},
  year = {2022},
  month = {Jun},
  publisher = {American Physical Society},
  doi = {10.1103/PhysRevB.105.245418},
  url = {https://link.aps.org/doi/10.1103/PhysRevB.105.245418}
}

@article{
CrossnoScience2016,
author = {Jesse Crossno  and Jing K. Shi  and Ke Wang  and Xiaomeng Liu  and Achim Harzheim  and Andrew Lucas  and Subir Sachdev  and Philip Kim  and Takashi Taniguchi  and Kenji Watanabe  and Thomas A. Ohki  and Kin Chung Fong },
title = {Observation of the Dirac fluid and the breakdown of the Wiedemann-Franz law in graphene},
journal = {Science},
volume = {351},
number = {6277},
pages = {1058-1061},
year = {2016},
doi = {10.1126/science.aad0343},
URL = {https://www.science.org/doi/abs/10.1126/science.aad0343}
}

@article{KiselevPRB2023,
  title = {Generalized Wiedemann-Franz law in a two-site charge Kondo circuit: Lorenz ratio as a manifestation of the orthogonality catastrophe},
  author = {Kiselev, M. N.},
  journal = {Phys. Rev. B},
  volume = {108},
  issue = {8},
  pages = {L081108},
  numpages = {6},
  year = {2023},
  month = {Aug},
  publisher = {American Physical Society},
  doi = {10.1103/PhysRevB.108.L081108},
  url = {https://link.aps.org/doi/10.1103/PhysRevB.108.L081108}
}

@article{LeePRL2020,
  title = {Fractional Mutual Statistics on Integer Quantum Hall Edges},
  author = {Lee, June-Young M. and Han, Cheolhee and Sim, H.-S.},
  journal = {Phys. Rev. Lett.},
  volume = {125},
  issue = {19},
  pages = {196802},
  numpages = {6},
  year = {2020},
  month = {Nov},
  publisher = {American Physical Society},
  doi = {10.1103/PhysRevLett.125.196802},
  url = {https://link.aps.org/doi/10.1103/PhysRevLett.125.196802}
}

@article{
IftikharScience2018,
author = {Z. Iftikhar  and A. Anthore  and A. K. Mitchell  and F. D. Parmentier  and U. Gennser  and A. Ouerghi  and A. Cavanna  and C. Mora  and P. Simon  and F. Pierre },
title = {Tunable quantum criticality and super-ballistic transport in a “charge” Kondo circuit},
journal = {Science},
volume = {360},
number = {6395},
pages = {1315-1320},
year = {2018},
doi = {10.1126/science.aan5592},
URL = {https://www.science.org/doi/abs/10.1126/science.aan5592}}

@Article{IftikharNature2015,
author={Iftikhar, Z.
and Jezouin, S.
and Anthore, A.
and Gennser, U.
and Parmentier, F. D.
and Cavanna, A.
and Pierre, F.},
title={Two-channel Kondo effect and renormalization flow with macroscopic quantum charge states},
journal={Nature},
year={2015},
month={Oct},
day={01},
volume={526},
number={7572},
pages={233-236},
issn={1476-4687},
doi={10.1038/nature15384},
url={https://doi.org/10.1038/nature15384}
}

@article{KarkiPRB2025,
  title = {Multimode Coulomb blockade oscillations},
  author = {Karki, D. B.},
  journal = {Phys. Rev. B},
  volume = {111},
  issue = {3},
  pages = {035426},
  numpages = {10},
  year = {2025},
  month = {Jan},
  publisher = {American Physical Society},
  doi = {10.1103/PhysRevB.111.035426},
  url = {https://link.aps.org/doi/10.1103/PhysRevB.111.035426}
}

@article{KarkiPRB2024,
  title = {Quantum criticality in coupled hybrid metal-semiconductor islands},
  author = {Karki, D. B.},
  journal = {Phys. Rev. B},
  volume = {110},
  issue = {24},
  pages = {L241406},
  numpages = {5},
  year = {2024},
  month = {Dec},
  publisher = {American Physical Society},
  doi = {10.1103/PhysRevB.110.L241406},
  url = {https://link.aps.org/doi/10.1103/PhysRevB.110.L241406}
}

@article{KarkiPRL2023,
  title = {${\mathbb{Z}}_{3}$ Parafermion in the Double Charge Kondo Model},
  author = {Karki, D. B. and Boulat, Edouard and Pouse, Winston and Goldhaber-Gordon, David and Mitchell, Andrew K. and Mora, Christophe},
  journal = {Phys. Rev. Lett.},
  volume = {130},
  issue = {14},
  pages = {146201},
  numpages = {7},
  year = {2023},
  month = {Apr},
  publisher = {American Physical Society},
  doi = {10.1103/PhysRevLett.130.146201},
  url = {https://link.aps.org/doi/10.1103/PhysRevLett.130.146201}
}

@misc{hurvitz2025metallicislandarraysynthetic,
      title={Metallic island array as synthetic quantum matter: fractionalized entropy and thermal transport}, 
      author={Nitay Hurvitz and Gleb Finkelstein and Eran Sela},
      year={2025},
      eprint={2510.20491},
      archivePrefix={arXiv},
      primaryClass={cond-mat.str-el},
      url={https://arxiv.org/abs/2510.20491}, 
}

@Article{HartmanNatPhys2018,
author={Hartman, Nikolaus
and Olsen, Christian
and L{\"u}scher, Silvia
and Samani, Mohammad
and Fallahi, Saeed
and Gardner, Geoffrey C.
and Manfra, Michael
and Folk, Joshua},
title={Direct entropy measurement in a mesoscopic quantum system},
journal={Nature Physics},
year={2018},
month={Nov},
day={01},
volume={14},
number={11},
pages={1083-1086},
issn={1745-2481},
doi={10.1038/s41567-018-0250-5},
url={https://doi.org/10.1038/s41567-018-0250-5}
}

@article{ZhangPRR2025,
  title = {Ballistic-to-diffusive transition in engineered counterpropagating quantum Hall channels},
  author = {Zhang, Aifei and Watanabe, Kenji and Taniguchi, Takashi and Roche, Patrice and Altimiras, Carles and Parmentier, Francois D. and Maillet, Olivier},
  journal = {Phys. Rev. Res.},
  volume = {7},
  issue = {4},
  pages = {L042037},
  numpages = {6},
  year = {2025},
  month = {Nov},
  publisher = {American Physical Society},
  doi = {10.1103/3d25-wpth},
  url = {https://link.aps.org/doi/10.1103/3d25-wpth}
}

@article{KanePRL1994,
  title = {Randomness at the edge: Theory of quantum Hall transport at filling \ensuremath{\nu}=2/3},
  author = {Kane, C. L. and Fisher, Matthew P. A. and Polchinski, J.},
  journal = {Phys. Rev. Lett.},
  volume = {72},
  issue = {26},
  pages = {4129--4132},
  numpages = {0},
  year = {1994},
  month = {Jun},
  publisher = {American Physical Society},
  doi = {10.1103/PhysRevLett.72.4129},
  url = {https://link.aps.org/doi/10.1103/PhysRevLett.72.4129}
}

\section{End matter}
The end matter provides  A/ further details on the derivations of of the non-interacting temperature profile Eq. (2) and B/  the general solution of the temperature profile in the strong interacting limit which approximates to Eq. (7) in the large size $M\gg\Delta$ limit.
\subsection{A - Temperature profile in the non-interacting/diffusive limit}
The temperature profile in the diffusive limit, Eq. (\ref{T_sq_noninteracting}), is obtained from Eq. (\ref{charge_conservation}). In the limit $G_0/C\ll k_\mathrm{B}T/\hbar$, current at the dominant frequencies $\lesssim k_\mathrm{B}T/\hbar$ flows through the capacitance, meaning that each island's potential at relevant frequencies is no longer floating, i.e. $\delta V_k(\omega)\approx0$. This means that for each channel, 
\begin{equation}
\label{approx_fluctu_JNonly}
    \delta I_{\gamma,\ell}^{k}=\delta I_{\gamma,\ell}^{T_k}+G_0\delta V_k\underset{G_0/C\ll k_\mathrm{B}T_k/\hbar}{\approx}\delta I_{\gamma,\ell}^{T_k} 
\end{equation}
As a result, in Eq. (\ref{charge_conservation}) we are only left with Johnson-Nyquist current noises for each channel, which are uncorrelated with each other. After multiplying the sum over $\ell$ by itself in frequency domain and taking the average, the  emitted power per channel is $P_{\gamma,\ell}^k=\pi^2k_\mathrm{B}^2T_k^2/6h$. From the heat balance at each island $k$, this leads to
\begin{equation}
    \label{Heat_diffusion_no_interactions}
    T_{k+1}^2-2T_k^2+T_{k-1}^2=0.
\end{equation}
This is nothing but the diffusion equation (\ref{Heat_diffusion}), without a second term originating from interactions. From there, we obtain a simple relation, $T_k^2-T_{k-1}^2=T_{k+1}^2-T_k^2$. This is easily solved and, accounting for the boundary conditions $T_{0(M+1)}=T_\mathrm{S(D)}^2$, we arrive at the non-interacting temperature profile of Eq. (\ref{T_sq_noninteracting}). This situation is, incidentally, a good illustration of the equivalence, in the non-interacting limit, of the "plasmon" Langevin approach used in the main text, and the Landauer-Büttiker approach where each island is a reservoir of non-interacting electrons. In the latter approach, the power accompanying electronic waves emitted in each channel $\ell$ of island $k$ writes, according to scattering theory:
\begin{equation}
\begin{split}
    \label{Power_LB}
    P_{\gamma,\ell}^k&=\frac{1}{h}\int_0^{+\infty}\mathrm{d}E\left(E-\mu\right)\left[f_k(E,\mu)-\Theta(\mu-E)\right]\\&=\frac{\pi^2k_\mathrm{B}^2T_k^2}{6h},
\end{split}
\end{equation}
with $f_k(E,\mu)=1/\left(\exp[(E-\mu)/k_\mathrm{B}T_k]+1\right)$ the Fermi-Dirac distribution for reservoir $k$. From there we use the same balance arguments to recover the profile of Eq. (\ref{T_sq_noninteracting}).
\subsection{B - General solution for the temperature profile in the interacting limit}
We write the reduced version of the diffusion equation (\ref{Heat_diffusion}), with $\Delta=\sqrt{(N-1)(M+1)}/2$ the decay length:
\begin{equation}
    \label{SI_heat_diff_reduced}
    \theta_{k+1}-\left(2+\frac{1}{\Delta^2}\right)\theta_k+\theta_{k-1}=0,
\end{equation}
with $\theta_k=T_k^2-T_\mathrm{mid}^2$. The two roots of the characteristic equation associated to Eq. (\ref{SI_heat_diff_reduced}) are:
\begin{equation}
    \label{SI_carac_roots}
    \lambda_{\pm}=1+\frac{1}{2\Delta^2}\pm\sqrt{\left(1+\frac{1}{2\Delta^2}\right)^2-1}.
\end{equation}
We then inject the ansatz $\theta_k=\alpha\lambda_+^k+\beta\lambda_-^k$ into Eq. (\ref{SI_heat_diff_reduced}) and use boundary conditions $\theta_{0/M+1}=T_\mathrm{S/D}^2-T_\mathrm{mid}^2$ to obtain the general solution for the squared temperature profile:
\begin{equation}
    \label{SI_Tsq_general}
    T_k^2=T_\mathrm{mid}^2+\frac{\lambda_-^k\left(1+\lambda_+^{M+1}\right)-\lambda_+^k\left(1+\lambda_-^{M+1}\right)}{\lambda_+^{M+1}-\lambda_-^{M+1}}\frac{\left(T_\mathrm{S}^2-T_\mathrm{D}^2\right)}{2}.
\end{equation}
In the long chain limit, $\Delta\gg1$, we have $\lambda_\pm\approx(1\pm1/\Delta)$, and we can make the following approximations: $\lambda_-^{M+1}\approx 0$, $\lambda_+^{M+1}=e^{(M+1)\log\lambda_+}\approx e^{(M+1)/\Delta}$, and $\lambda_\pm^k=e^{k\log\lambda_\pm}\approx e^{\pm k/\Delta}$. Then Eq. (\ref{SI_Tsq_general}) simplifies and we obtain the approximate profile of temperature (\ref{T_profile_simple}).

\newpage
\onecolumngrid
\setcounter{figure}{0}
\renewcommand{\thepage}{S\arabic{page}} 
\renewcommand{\thesection}{S\arabic{section}}  
\renewcommand{\thetable}{S\arabic{table}}  
\renewcommand{\thefigure}{S\arabic{figure}} 
\renewcommand{\theequation}{S.\arabic{equation}}
\section*{Supplementary Material for ``Ballistic-to-diffusive transition in engineered counter-propagating quantum Hall channels"}
In this supplementary material we detail some calculations that are not shown in the main text, as well as the decomposition in neutral and charged modes described in the main text. We also show an alternative derivation for power emission.
\section{Writing voltages as functions of currents in the physical basis}
We recall the current conservation equation written in the main text, for each island $k$, using islands' charges related to voltages by $\delta V_k=q_k/C$:
\begin{equation}
    \label{SI_current_conservation}
    \frac{\mathrm{d}q_k}{\mathrm{d}t}=\sum_{\ell=1}^N\left(\delta I^{T_{k-1}}_{\mathrm{R},\ell}+\delta I^{T_{k+1}}_{\mathrm{L},\ell}-\delta I^{T_k}_{\mathrm{R},\ell}-\delta I^{T_k}_{\mathrm{L},\ell}\right)+N\omega_c(q_{k-1}-2q_k+q_{k+1}),
\end{equation}
with $k$ the island index that runs from 1 to $M$, $k=0$ ($M+1$) being the source with $\delta I_{\mathrm{R},\ell}^{T_0}=\delta I^{T_\mathrm{S}}_\ell$ (drain with $\delta I_{\mathrm{L},\ell}^{T_{M+1}}=\delta I^{T_\mathrm{D}}_\ell$) and $\omega_c=G_0/C$ the cut-off frequency above which current may flow through the capacitors. Condensing the current notation $\mathcal{I}_k=\sum_{\ell=1}^N\left(\delta I^{T_{k-1}}_{\mathrm{R},\ell}+\delta I^{T_{k+1}}_{\mathrm{L},\ell}-\delta I^{T_k}_{\mathrm{R},\ell}-\delta I^{T_k}_{\mathrm{L},\ell}\right)$ and writing in Fourier space, we have:
\begin{equation}
\label{SI_current_conservation_Fourier}
    \mathcal{I}_k(\omega)=-N\omega_c\left[q_{k-1}(\omega)+q_{k+1}(\omega)\right]+\chi^{-1}(\omega)q_k(\omega),
\end{equation}

where we have introduced the charge susceptibility $\chi(\omega)=(i\omega+2N\omega_c)^{-1}$. Eq. (\ref{SI_current_conservation_Fourier}) can be condensed in matrix form: $\textbf{J}(\omega)=\mathcal{Y}(\omega)\textbf{q}(\omega)$, with $\textbf{J}=(\mathcal{I}_1,\dots,\mathcal{I}_M)$, $\textbf{q}=(q_1,\dots,q_M)$ and matrix

\begin{equation}
    \label{SI_trig_admittance_matrix}
\mathcal{Y}(\omega)=
\begin{bmatrix}
\chi^{-1}(\omega) & -N\omega_c & 0 & \cdots & 0 \\
-N\omega_c & \ddots & \ddots & \ddots & \vdots \\
0 & \ddots & \ddots & \ddots & 0 \\
\vdots & \ddots & \ddots & \ddots & -N\omega_c \\
0 & \cdots & 0 & -N\omega_c & \chi^{-1}(\omega)
\end{bmatrix}.
\end{equation}

Matrix $\mathcal{Y}$ is tri-diagonal and has a generic yet cumbersome invertible form. We choose, for the sake of simplicity, to focus on the HCB regime, where $N\hbar\omega_c\gg k_\mathrm{B}T$. In that limit, current thermal fluctuations, which are cut off above frequency $k_\mathrm{B}T/\hbar$, cannot flow through the capacitors, and $\chi^{-1}\approx2N\omega_c$. The inverse matrix of $\mathcal{Y}$ then has a simple enough form to provide the following analytical expressions for each island's voltage $\delta V_k$:

\begin{equation}
    \delta V_k(\omega)\underset{k_\mathrm{B}T\ll\hbar\omega_c}{\approx}\frac{M+1-k}{G_0N(M+1)}\sum_{j=1}^kj\mathcal{I}_j(\omega)+\frac{k}{G_0N(M+1)}\sum_{j=k+1}^M(M+1-j)\mathcal{I}_j(\omega),
\end{equation}

leading after rearranging the sums to the expression shown in the main text.
\section{Decomposition in charged and neutral chiral bosonic fields}
Here we briefly recall the context and provide a brief mathematical background for the decomposition in charged and neutral modes (unidirectional and split) that is qualitatively described in the main text, focusing on one island. For a more detailed description of chiral edge channels in terms of bosonic charge fields, in presence of an Ohmic contact, see e.g. \cite{SlobodeniukPRB2013,MorelPRB2024}. We introduce chiral bosonic fields $\phi_\mathrm{L/R,\ell,+/-}$, associated to physical incoming/outgoing ($-/+$) edge channels, according to $\rho_\mathrm{L/R,\ell,+/-}=\mp\partial_x\phi_\mathrm{L/R,\ell,+/-}/2\pi$, where $\rho_\mathrm{L/R,\ell,+/-}$ are the charge densities of each channel. Bosonic fields are related to currents via $\delta I=-e\partial_t\phi/2\pi$ and obey the commutation relation $\left[\partial_x\phi_{\gamma,\ell,\sigma}(x,t),\phi_{\gamma',\ell',\sigma'}(x',t)\right]=2\pi i\sigma\delta_{\gamma,\gamma'}\delta_{\ell,\ell'}\delta(x-x')$. The low-energy, effective Hamiltonian for the Ohmic contact interfaced with $2N$ incoming and $2N$ outgoing edge channels writes:
\begin{equation}
    \label{SI_Hamiltonian_island}
    \mathcal{H}=\frac{\hbar v}{4\pi}\sum_{\gamma,\ell,\sigma}\int_{-\infty}^{+\infty}\mathrm{d}x\left(\frac{\partial\phi_{\gamma,\ell,\sigma}}{\partial x}\right)^2+\frac{\mathcal{Q}^2}{2C},
\end{equation}
where the $x<0$ region corresponds to the island (with channels that decay over very short distances) and $x>0$ to the QH channels (that are for simplicity all put on the same side here, without loss of generality since we treat a single island). Here the second term in the Hamiltonian corresponds to the island's charging energy and acts as a local interaction term, with a charge operator:
\begin{equation}
    \label{SI_Charge_operator}
    \mathcal{Q}=\sum_{\gamma,\ell,\sigma}\int_{-\infty}^0\rho_{\gamma,\ell,\sigma}(x,t)\mathrm{d}x=\sum_{\gamma,\ell}\left[\phi_{\gamma,\ell,-}(0)-\phi_{\gamma,\ell,+}(0)\right]
\end{equation}

Note that this enables, by combination of commutation relations with Heisenberg's equations of motion applied to each island $k$ in the chain configuration, to recover the charge conservation equations \ref{SI_current_conservation}, noting that incoming fields to island $k$ are the outgoing right/left fields of island $k-1/k+1$ . 

We now focus on the alternative representation of bosonic fields. We focus on outgoing modes and drop the $+/-$ subscript for convenience, but the decomposition shown here equally applies to incoming ones. Following Ref. \cite{MorelPRB2024}, we construct new eigenmodes that are particularly suited for the HCB regime and diagonalize the Hamiltonian \ref{SI_Hamiltonian_island}. They decompose as $2N-1$ neutral modes, which do not carry a net charge and therefore do not couple \textit{directly} to the island's charge degree of freedom. The one that carries all the net charge, the C charged mode, writes:

\begin{equation}
    \label{SI_charge_mode}
    \tilde{\phi}_c=\frac{1}{\sqrt{2N}}\sum_{\ell=1}^N\left(\phi_{\mathrm{L},\ell}+\phi_{\mathrm{R},\ell}\right).
\end{equation}

We now turn to the neutral modes. Among these, we single out the split neutral (SN) mode, which provides net charges to adjacent islands that trigger charge modes in cascade:

\begin{equation}
    \label{SI_split_neutral_mode}
    \tilde{\phi}_s=\frac{1}{\sqrt{2N}}\sum_{\ell=1}^N\left(\phi_{\mathrm{L},\ell}-\phi_{\mathrm{R},\ell}\right).
\end{equation}

The remaining modes are directional neutral (LN/RN) modes, i.e., they are emitted in one direction only:

\begin{equation}
    \label{SI_directional_neutral_mode}
    \tilde{\phi}_{\gamma,m}=\sqrt{\frac{N-m}{N-m+1}}\left(\phi_{\gamma,m}-\frac{1}{N-m}\sum_{\ell=m+1}^N\phi_{\gamma,\ell}\right).
\end{equation}

The matrix $\mathcal{O}$ such that $\tilde{\Phi}=\mathcal{O}\Phi$, with $\Phi=(\{\phi_{\gamma,\ell}\})$ and $\tilde{\Phi}=(\tilde{\phi}_s,\tilde{\phi}_c,\{\tilde{\phi}_{\gamma,m}\})$ satisfies $\mathcal{O}\,^\mathrm{T}\mathcal{O}=1$. Thus, the new bosonic modes that are referred to in the main text do form an orthonormal basis.

\section{Derivation of emitted power using the charged and neutral modes basis}
\begin{figure}[ht!]
    \centering
    \includegraphics[width=0.5\linewidth]{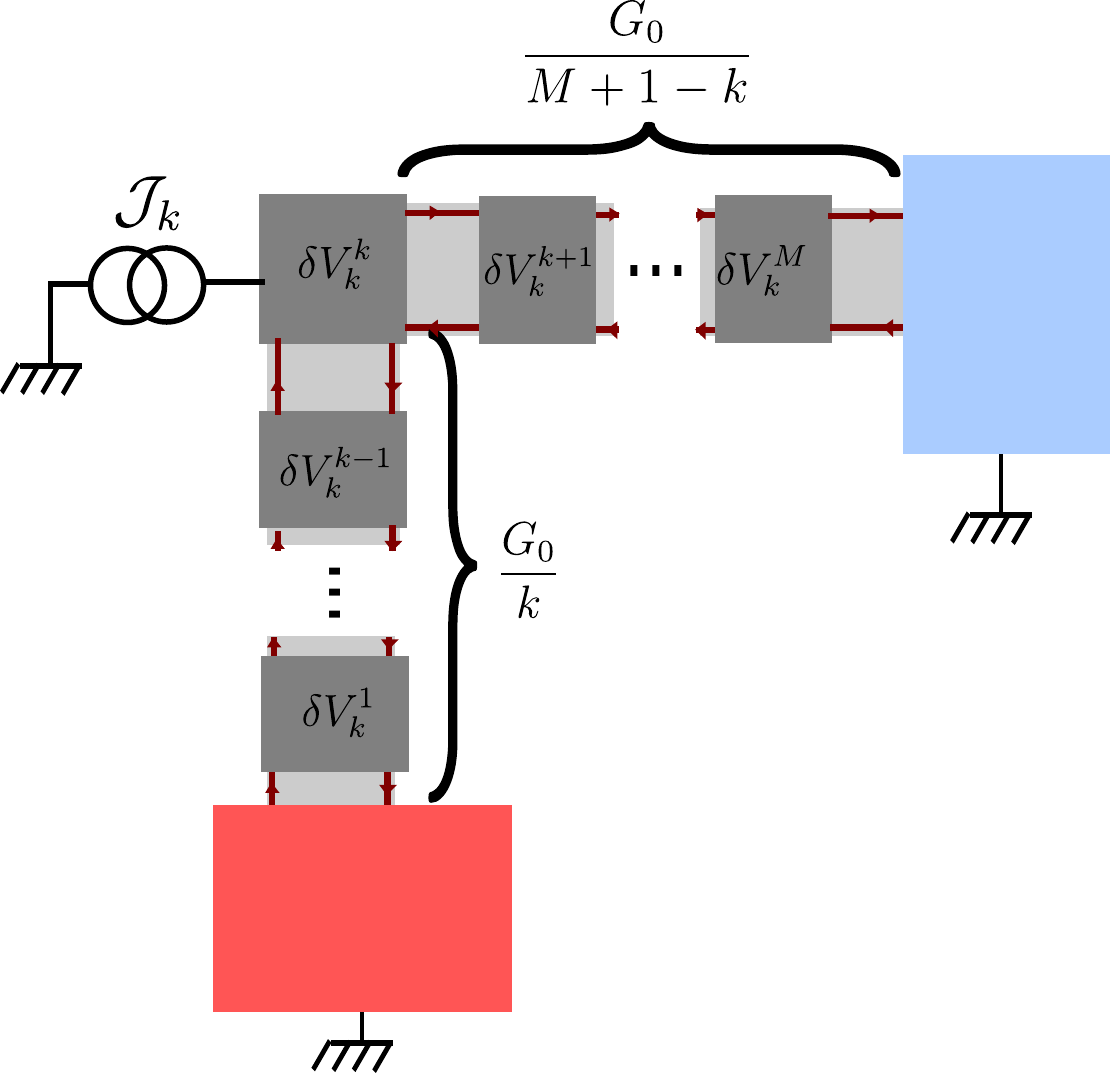}
    \caption{Equivalent representation for voltage build-up, assuming net thermal current noise on each island behaves as an ideal current source.}
    \label{fig:HCB}
\end{figure}
Here we present an alternative derivation for the power emitted by an island, using the description of charged and neutral modes. For simplicity we introduce two charge fields $\tilde{\phi}_\mathrm{c,L}$ and $\tilde{\phi}_\mathrm{c,R}$ that correspond to net current emission on the left and right, respectively, of the island. They are constructed simply from the sum and difference of the C field $\tilde{\phi}_c$ and the SN field $\tilde{\phi}_s$ introduced earlier, and are bound to cancel each other in the HCB regime.

With this decomposition, the remaining $2N-2$ neutral modes do not couple to the islands' charges and are just emitted and absorbed by successive islands. The power emitted in one direction by neutral modes (say, by the $N-1$ LN modes on the left) is then just $(N-1)\kappa_0T_k^2$. All the current is then carried by the two charge modes on left and right. We directly use the semi-classical Langevin approach and associate Johnson-Nyquist current fluctuations $\delta I_\mathrm{L/R}^{T_k}$ to the two charge modes $\tilde{\phi}_\mathrm{c,L/R}$. The idea is to consider the net Johnson-Nyquist current quantity $\mathcal{J}_k=\delta I_\mathrm{L}^{T_{k+1}}+\delta I_\mathrm{R}^{T_{k-1}}-\delta I_\mathrm{L}^{T_{k}}-\delta I_\mathrm{R}^{T_{k}}$ impinging on an island as a current source, that is essentially DC with respect to the chain's $RC$ cutoff frequency, and feeds two resistors made each of one side of the chain towards source or drain, with respective values $k/G_0$ and $(M+1-k)/G_0$. The resulting build-up voltage for each island, under this assumption, can be readily obtained from Landauer-Büttiker formalism. For an island $j\leq k$, the voltage build-up caused by the current source of island $k$ is:
\begin{equation}
    \label{SI_partial_volt_buildup_left}
    \delta V_j^k=\frac{j}{k}\frac{\mathcal{J}_k}{\frac{G_0}{M+1-k}+\frac{G_0}{k}}=\frac{j(M+1-k)\mathcal{J}_k}{(M+1)G_0},
\end{equation}
while for an island $j>k$ it writes:
\begin{equation}
    \label{SI_partial_volt_buildup_rightt}
    \delta V_j^k=\frac{M+1-j}{M+1-k}\frac{\mathcal{J}_k}{\frac{G_0}{M+1-k}+\frac{G_0}{k}}=\frac{k(M+1-j)\mathcal{J}_k}{(M+1)G_0}.
\end{equation}
Summing then all contributions to voltage build-up in island $j$ from sources on islands $k$, we obtain:
\begin{equation}
    \label{SI_deltaV_buildup_1}
    \delta V_j =\sum_{k=1}^N\delta V_j^k=(M+1-j)\sum_{k=1}^{k\leq j}\frac{k}{G_0(M+1)}\mathcal{J}_k+j\sum_{k=1}^{k>j}\frac{M+1-k}{G_0(M+1)}\mathcal{J}_k.
\end{equation}
After rearranging the sums, we have:
\begin{equation}
    \label{SI_deltaV_buildup_2}
    \delta V_j =\frac{M+1-j}{G_0(M+1)}\left[\delta I^{T_\mathrm{S}}_\mathrm{R}-\delta I_\mathrm{L}^{T_j}+\sum_{k=1}^{j-1}\left(\delta I_\mathrm{R}^{T_k}-\delta I_\mathrm{L}^{T_k}\right)\right]+\frac{j}{G_0(M+1)}\left[\delta I^{T_\mathrm{D}}_\mathrm{L}-\delta I_\mathrm{R}^{T_j}+\sum_{k=j+1}^{M}\left(\delta I_\mathrm{L}^{T_k}-\delta I_\mathrm{R}^{T_k}\right)\right].
\end{equation}
From this, we can write the power emitted through each charge mode: $P^k_\mathrm{c,L/R}=\int_{-\infty}^{+\infty}S_{I^{k}_\mathrm{L/R}}(\omega)\mathrm{d}\omega/4\pi G_0$, with $2\pi S_{I^{k}_\mathrm{L/R}}(\omega)\delta(\omega+\omega')=\left\langle\delta I_\mathrm{L/R}^k(\omega)\delta I_\mathrm{L/R}^k
(\omega')\right\rangle$. The correlation function writes:
\begin{equation}
        \label{SI_power_charge_mode}
\left\langle\delta I_\mathrm{L/R}^k(\omega)\delta I_\mathrm{L/R}^k
(-\omega)\right\rangle=\left\langle\delta I_\mathrm{L/R}^{T_k}(\omega)\delta I_\mathrm{L/R}^{T_k}
(-\omega)\right\rangle+2\left\langle\delta I_\mathrm{L/R}^{T_k}(\omega)\delta V_k
(-\omega)\right\rangle+\left\langle\delta V_k
(\omega)\delta V_k
(-\omega)\right\rangle.
\end{equation}
After injection of Eq.(\ref{SI_deltaV_buildup_2}) and the integration of spectra, we obtain for the left power emission:
\begin{equation}
    \label{SI_Pleft_charged}
    P^k_\mathrm{c,L}=\frac{2k^2}{(M+1)^2}T_k^2+\frac{(M+1-k)^2T_\mathrm{S}^2+k^2T_\mathrm{D}^2}{(M+1)^2}+2\frac{(M+1-k)^2\sum_{j=1}^{k-1}T_j^2+k^2\sum_{j=k+1}^MT_j^2}{(M+1)^2},
\end{equation}
while on the right side:
\begin{equation}
    \label{SI_Pright_charged}
    P^k_\mathrm{c,R}=P^k_\mathrm{c,L}+\frac{2(M+1-2k)}{M+1}\kappa_0T_k^2.
\end{equation}
When adding the contribution $(N-1)\kappa_0T_k^2$ from the LN/RN neutral modes, we recover the expressions for power emission written in the main text.
\section{General solution for the temperature profile}
\subsection{Mean temperature definition}
Before showing the general profile of temperature, we first note that the diffusion equation obtained from the continuity of the heat flow initially writes:
\begin{equation}
    \label{SI_diffusion_equation_general}
(N-1)(T_{k+1}^2-2T_k^2+T_{k-1}^2)=\frac{4\left[(M+1)T_k^2+T_\mathrm{mid}^2-\sum_{j=0}^{M+1}T_j^2\right]}{(M+1)^2},
\end{equation}
and that, in the main text, we have shown a simplified version by making the identification $T_\mathrm{mid}^2=\sum_{j=0}^{M+1}T_j^2/(M+2)$. This can be justified by symmetry arguments. Indeed, the system is symmetric under temperature gradient reversal, i.e. by swapping source and drain temperatures, which are just the boundary conditions. Let us label $T_k'^2$ the solutions of the diffusion equation under gradient reversal, with $T_0'^2=T_\mathrm{D}^2$ and $T_{M+1}'^2=T_\mathrm{S}^2$. They must satisfy, by symmetry, $T_k'^2=T_{M+1-k}^2$. By linearity of the diffusion equation, $\{T_k^2+T_k'^2\}$ is the solution of the heat diffusion equation with boundary conditions $T_\mathrm{0}^2=T_\mathrm{M+1}^2=T_\mathrm{S}^2+T_\mathrm{D}^2=2T_\mathrm{mid}^2$, that is, for each $k$:
\begin{equation}
    \label{SI_sum_solution}
    \frac{T_k^2+T_{M+1-k}^2}{2}=T_\mathrm{mid}^2,
\end{equation}
or, rewritten differently:
\begin{equation}
    \label{SI_temperature_asymmetric}
    T_{M+1-k}^2-T_\mathrm{mid}^2=-\left(T_{k}^2-T_\mathrm{mid}^2\right)
\end{equation}
It follows that the squared temperature profile is anti-symmetric with respect to $T_\mathrm{mid}^2$ and to the middle of the chain, as visible in Fig. 2 of the main text, and from Eq. \ref{SI_temperature_asymmetric}, we have $\sum_{j=0}^{M+1}T_j^2=(M+2)T_\mathrm{mid}^2$.


\end{document}